\documentclass[aps, secnumarabic, nobibnotes, twocolumn, superscriptaddress]{revtex4-2}

\usepackage{amsfonts}
\usepackage{mathrsfs}
\usepackage{amsmath}
\usepackage{color}
\usepackage{natbib}
\usepackage{graphicx}
\usepackage{bm}
\usepackage{amssymb}
\usepackage{xspace}
\usepackage{epstopdf}
\usepackage{dcolumn}
\usepackage{longtable}
\usepackage{multirow}
\usepackage[title,titletoc]{appendix}
\usepackage{graphicx}
\usepackage{float}
\usepackage[colorlinks=true, letterpaper=true, pdfstartview=FitV, linkcolor=blue, citecolor=blue, urlcolor=blue]{hyperref}

\makeatletter

\newcommand{\Rmnum}[1]{\expandafter\@slowromancap\romannumeral #1@}
\makeatother
\begin{document}


\title{Controllable Weyl Nodes and Fermi Arcs from Floquet Engineering Triple Fermions}%

\author{Shengpu Huang}
\affiliation{Institute for Structure and Function $\&$ Department of Physics $\&$ Chongqing Key Laboratory for Strongly Coupled Physics, Chongqing University, Chongqing 400044, People's Republic of China}

\author{Fangyang Zhan}
\affiliation{Institute for Structure and Function $\&$ Department of Physics $\&$ Chongqing Key Laboratory for Strongly Coupled Physics, Chongqing University, Chongqing 400044, People's Republic of China}

\author{Xianyong Ding}
\affiliation{Institute for Structure and Function $\&$ Department of Physics $\&$ Chongqing Key Laboratory for Strongly Coupled Physics, Chongqing University, Chongqing 400044, People's Republic of China}

\author{Dong-Hui Xu}
\email{donghuixu@cqu.edu.cn}
\affiliation{Institute for Structure and Function $\&$ Department of Physics $\&$ Chongqing Key Laboratory for Strongly Coupled Physics, Chongqing University, Chongqing 400044, People's Republic of China}

\author{Da-Shuai Ma}
\email{madason.xin@gmail.com}
\affiliation{Institute for Structure and Function $\&$ Department of Physics $\&$ Chongqing Key Laboratory for Strongly Coupled Physics, Chongqing University, Chongqing 400044, People's Republic of China}
\affiliation{Center of Quantum materials and devices, Chongqing University, Chongqing 400044, People's Republic of China}

\author{Rui Wang}
\email[]{rcwang@cqu.edu.cn}
\affiliation{Institute for Structure and Function $\&$ Department of Physics $\&$ Chongqing Key Laboratory for Strongly Coupled Physics, Chongqing University, Chongqing 400044, People's Republic of China}
\affiliation{Center of Quantum materials and devices, Chongqing University, Chongqing 400044, People's Republic of China}


\begin{abstract}
 Floquet engineering with periodic driving as a powerful tool for designing desirable topological states has been the subject of intense recent studies.
 Here, we present the application of Floquet engineering to investigate evolution of topological triple fermions under irradiation of circularly polarized light (CPL), a phenomenon that currently remains a mystery. By using first-principles calculations and Floquet theorem, we demonstrate that WC-type TiO and its analogues are promising candidates for Floquet engineering of triple fermions. The symmetry analysis reveals that the electric field of CPL can break the specific symmetries, such as the time-reversal symmetry and its combination of spatial symmetries, inducing a transition to a flexibly controllable Weyl semimetallic phase. The survived spatial symmetry, controlled by light, guarantees that the Weyl nodes are located along the high-symmetry line or in high-symmetry planes in momentum space. Our findings focusing on Floquet engineering in realistic materials featured by triple fermions would facilitate both theoretical and experimental interest.
\end{abstract}
\maketitle


\textit{\textcolor{blue}{Introduction. ---}}
In recent decades, there has been a growing interest in the investigation of intriguing quantum phenomena associated with nontrivial band topology~\cite{2016RevModPhys.88.021004,2019RevModPhys.91.015005,maTopologicalPhasesAcoustic2019}.
The developments of various topological phases and their material realization emphasize the importance of symmetry, such as $\mathbb{Z}_2$ topological insulators~\cite{2006doi:10.1126/science.1133734,2008hsieh2008topological,2010RevModPhys.82.3045}, quantized electric multipole insulators~\cite{benalcazar2017quantized,2017PhysRevB.96.245115,schindler2018higher}, and topological semimetals (TSMs)~\cite{burkov2016topological,yan2017topological,2021RevModPhys.93.025002}.
Distinct from gapped topological phases, TSMs possessing the gapless nodal points (or nodal lines) near the Fermi level are particularly relevant to symmetries. The positions of nodal points (or nodal lines) are determined by specific symmetries in momentum space, and quasiparticles characterized by topological invariants exhibit distinct band degeneracy \cite{2020PhysRevB.101.235119,PhysRevB.106.214309,2021PhysRevB.103.L161109}, dispersion types\cite{soluyanov2015type,2021li2021type}, and dimensions\cite{2011PhysRevB.84.235126,2015PhysRevB.92.045108}.
The nontrivial band topology of TSMs often leads to attractive phenomena, sucn as ultrahigh carrier mobility~\cite{shekhar2015extremely,liang2015ultrahigh}, half-integer quantum Hall effects~\cite{novoselov2005two,zhang2005experimental}, large diamagnetism~\cite{ali2014large,chen2016extremely,kumar2017extremely}, and electromagnetic duality~\cite{2017PhysRevB.95.205108,2022PhysRevLett.128.027201}, and thus considered to have a wide range of applications in future devices and technologies.
Since the realization of various topological quasiparticles in TSMs strongly depends on symmetries, the studies of various symmetry-protected TSM phases and their evolution by applying symmetry-breaking perturbations have attracted intense studies, which is made explicit in topological classification.

Recently, Floquet engineering with periodic driving offers a powerful tool for engineering quantum states, and thus has been extensively employed to study various quantum systems with intriguing topological features \cite{2013PhysRevLett.110.026603,2018PhysRevLett.120.237403,2018PhysRevB.84.235108,2018PhysRevLett.120.156406,zhou2023pseudospin,zhan2023floquet,2023PhysRevB.107.085151,mciver2020light,PhysRevB.79.081406}.
Due to the naturally breaking of time-reversal symmetry, circularly polarized light (CPL) was verified experimentally to bring the Dirac mass term to the Dirac cones of graphene or surface states of topological insulators Bi$_2$Se$_3$, resulting in Chern insulators~\cite{PhysRevB.79.081406} or Floquet-Bloch states~\cite{wang2013observation,add1merboldt2024observationfloquetstatesgraphene}.
Benefiting the efforts donated to uncover mappings between symmetries and band topology~\cite{bradlyn2016beyond,yu2022encyclopedia,2022PhysRevB.105.104426,2023PhysRevB.107.075405,add3tang2024group,add2}, it is now possible to actively manipulate the transition between different topological states and thereby design desired TSMs under the irradiation of light fields.
For instance, it has shown that Floquet Weyl semimetallic phases can be generated in topological insulators, Dirac semimetals, and nodal-line semimetals subjected to periodic driving light fields~\cite{2016PhysRevLett.117.087402,hubener2017creating,2020flo_deng_deng2020photoinduced}.
Very recently, bicircular light was demonstrated to offer a versatile way to control magnetic symmetries and lead to a noncentrosymmetric magnetic Weyl semimetallic phase from the Dirac semimetal Cd$_3$As$_2$~\cite{PRL2022BCL}.
To date, the theoretical investigation of light-driven TSM phases and their transition has predominantly relied on employing low-energy effective models; however the central task lies in applying Floquet engineering to realistic materials, which is still in its infancy.
Moreover, Floquet engineering of band evolution and topological transition in TSMs with triple fermions (i.e., low-energy quasiparticles near the triply degenerate nodal points and being considered as an intermediate state between Dirac and Weyl states~\cite{2016PhysRevX.6.031003,lv2017observation,ma2018three}) remains unexplored.

\begin{figure}[t]
\includegraphics[width=\linewidth]{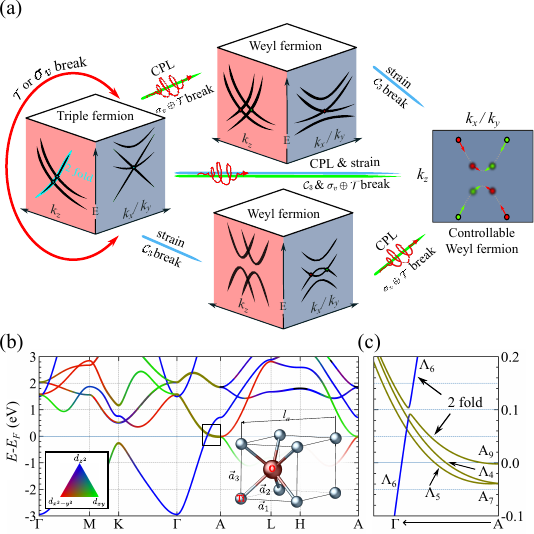}
\caption{\label{fig:fig1}
(a) The sketched figure illustrated the external field and corresponding symmetry breaking induced topological transition in this work. The dispersion along $k_z$ and $k_x/k_y$ for each phase are illustrated, with the dots in red and green mark different chiralities.
(b) The electronic band structures of intrinsic WC-type TiO along high-symmetry lines SOC.
(c) The enlarged view of band crossings along $\Gamma$-A, with irreducible representations of the high-symmetry line and A point are inserted.
Insert in (b) depicted the crystal structure of WC-type TiO. $\vec{a}_{1,2,3}$  is the lattice vectors of a hexagonal structure with $P\overline{6}m2$ ($D_{3h}$, No.~187). Here, the balls in brown and gray represent the O and Ti atoms, respectively. A uniaxial strain is applied along the direction represented by $l_a$.
}
\end{figure}

Here, we present that Floquet engineering on the evolution of topological triple fermions irradiated by periodic driving of CPL.
The key idea is to employ CPL irradiation to induce specific symmetry breaking, including the time-reversal symmetry and its combination of spatial symmetries, which directly impacts on the symmetry-protected triply degenerate nodal points [see Fig.~\ref{fig:fig1}(a)].
Based on first-principles calculations and Floquet theorem, we investigate light-dressed electronic properties and accompanied topological transition of WC-type TiO subject to CPL.
The WC-type TiO crystallizing in tungsten carbide type (WC-type) structure with a space group $P\overline{6}m2$ ($D_{3h}$, No.~187) was demonstrated to be a ideal candidate with triple fermions near the Fermi level ~\cite{ullah2023first,jin2023strain}.
As schematically illustrated in Fig.~\ref{fig:fig1}(a), the triply degenerate nodal points in WC-type TiO without light irradiation are protected by the threefold rotational symmetry $\mathcal{C}_3$, vertical mirror symmetry plus time-reversal symmetry $\mathcal{\sigma}_{v}\oplus\mathcal{T}$.
Under time-periodic and space-homogeneous CPL, we show that light irradiation can break the symmetry of $\mathcal{\sigma}_{v}\oplus\mathcal{T}$ and thereby lead to the triply degenerate nodal points splitting to twofold degenerate Weyl nodes. 
By carrying out Floquet engineering of strained WC-type TiO, the symmetry-breaking perturbations that couple of periodic-driving CPL with lattice strain can generate intriguing Weyl semimetallic phases with highly controllable Weyl nodes. More strikingly, we demonstrate that the survived spatial symmetry, controlled by light, guarantees that the Weyl nodes are located along the high-symmetry line or in high-symmetry planes in momentum space. Considering that the band topology of WC-type TiO has been well-studied \cite{ullah2023first,jin2023strain}, our findings would contribute to the advancement of Floquet engineering in investigating topological triple fermions, thereby facilitating both theoretical and experimental interest.

\textit{\textcolor{blue}{Intrinsic band topology of WC-type TiO. ---}}
We first revisit the intrinsic band topology of WC-type TiO.
Using first-principles calculations, we map the band structures of WC-type TiO and confirm that the triply degenerate nodal points exist in the presence of spin-orbit coupling (SOC).
In the main text, we only focus on the calculated electronic properties within SOC, and the results in the absence of SOC are included in the Supplemental Material (SM)~\cite{SM}.
The orbital-resolved band structures along high-symmetry paths are shown in Fig. \ref{fig:fig1}(b).
It seems that there is a band crossing point along the high-symmetry $\Gamma$-A path, which is mainly contributed by the $e_g$ orbital ($d_{xy}$ and $d_{x^2-y^2}$) and the $d_{z^2}$ orbital.
In fact, the enlarged view of the bands along $\Gamma$-A [see Fig.~\ref{fig:fig1}(c)] exhibits more complicated behaviors.
There are three sets of bands with distinct band degeneracy, i.e., the non-degenerate $\Lambda _{4}$ and $\Lambda _{5}$ bands, and the doubly degenerate $\Lambda _{6}$ band.
The $\Lambda _{4}(\Lambda _{5})$ band crosses with the doubly degenerate $\Lambda _{6}$ band, forming the triply degenerate nodal points that are protected by $\mathcal{C}_3$ and $\mathcal{\sigma}_{v}\oplus\mathcal{T}$, agreeing well with the previous works~\cite{2016PhysRevX.6.031003,2016PhysRevB.93.241202,2016PhysRevB.94.165201,ullah2023first,jin2023strain}.

\textit{\textcolor{blue}{Floquet engineering of pristine WC-type TiO. ---}}
To reveal the evolution of band topology in WC-type TiO by applying symmetry-breaking perturbations via Floquet engineering, we construct a symmetry-allowed tight-binding (TB) model in the magnetic space group $P\textbf{-}{6}m21'$ (No. 187.210).
Following the results of first-principles calculations, we consider the nearest and next nearest hopping, and the TB Hamiltonian under the spinful basis of $\left \{ \left | d_{z2} \uparrow  \right \rangle,\left | d_{z2} \downarrow  \right \rangle,\left | d_{xy} \uparrow \right \rangle,\left | d_{xy} \downarrow \right \rangle,\left | d_{x^2-y^2} \uparrow \right \rangle,\left | d_{x^2-y^2} \downarrow \right \rangle\right \}$ can be written as
 \begin{eqnarray}\label{Hamiltonian1}
H(\mathit{k})=
\begin{pmatrix}
f(\mathit{k})  &l(\mathit{k})^\ast   &n(\mathit{k})^\ast   \\
l(\mathit{k})  &g(\mathit{k})  &m(\mathit{k})^\ast  \\
n(\mathit{k})  &m(\mathit{k})  &h(\mathit{k})
\end{pmatrix}
\otimes \sigma _{0}+H_{\text{soc}}(\mathit{k}),
\end{eqnarray}
in which $\sigma _{0}$ represents the $2\times2$ unit matrix and $H_{\text{soc}}(\mathit{k})$ denotes the SOC effect, and the details of matrix elements of Hamiltonian Eq.(\ref{Hamiltonian1}) are summarized in the SM~\cite{SM}.
As shown in Fig. S1~\cite{SM}, it is found that the band topology of WC-type TiO near the Fermi level obtained from the effective TB Hamiltonian matches well with that from first-principles calculations.

\begin{figure}[t]
\includegraphics[width=\linewidth]{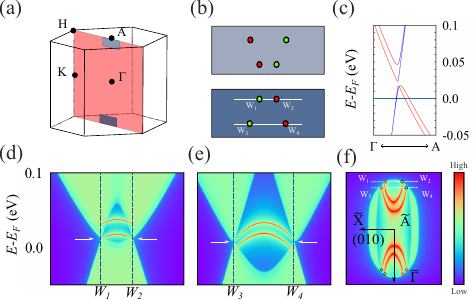}
\caption{\label{fig:fig3}
Intrinsic WC-type TiO under CPL light field.
(a) The electronic band structure along $\Gamma$-A under the light intensity of $eA_0/\hbar = 0.100$ \AA $^{-1}$. The enlarged views near Weyl nodes $W_b$ and $W_a$ are shown in (b)-(c).
(d)-(e) Surface spectrum of (010) surface along Weyl nodes $W_b$ and $W_a$.
The projected Weyl node points with chirality $+1/-1$ are highlighted by red/green circles.}
\end{figure}

Without light irradiation, the triply degenerate nodal points are protected by the coexistence of $\mathcal{C}_3$ and $\mathcal{\sigma}_{v}\oplus\mathcal{T}$. It is well known that CPL irradiation naturally breaks $\mathcal{T}$, which has been verified to serve as an effective means to achieve magnetic topological phases such as the quantum anomalous Hall effect~\cite{2018PhysRevLett.120.156406,2023PhysRevB.107.085151,zhan2023floquet,li2023tunable} and Weyl semimetal phases~\cite{2024PhysRevResearch.6.L012027}.
It is worth noting that, as a pseudovector, CPL radiation can also break the vertical mirror symmetry $\mathcal{\sigma}_{v}$.
Benefiting from this feature of CPL, in this work, we will see that CPL is an effective tool for driving topological transition of triple fermion.
We apply a counterclockwise CPL with a time-dependent vector potential $\mathbf{A}(t)=\mathrm{A}_{0}\left[\mathrm{cos}\left(\omega t\right),\mathrm{sin}\left(\omega t\right),0\right]$ to WC-type TiO, where $\mathrm{A}_{0}$ and $\omega$ correspond to the amplitude and the frequency of CPL.
The propagating direction is along the [001] direction and the polarized plane is parallel to the $x$-$y$ plane.
To avoid the Floquet sub-bands overlap with each other, the photon energy is set as $\hbar \omega= 16$~$\text{eV}$, which is larger than the band width of WC-type TiO. 
For the case of low-frequency light, the Floquet sub-bands cross over with each other resulting in complex band dispersion that is briefly discussed in SM~\cite{SM}.
As shown in Figs.~\ref{fig:fig3}(a) and \ref{fig:fig3}(b), we plot the band structures with the light intensity $e \mathrm{A}_0/\hbar = 0.100$ \AA $^{-1}$ along $\Gamma$-$A$.
We can see that light irradiation lifts the doubly degenerate band, forming two non-degenerate bands that are respectively featured by two $\mathcal{C}_3$ eigenvalues of $e^{i\pi/3}$ and $e^{-i\pi/3}$.
This phenomenon is similar to the case of applying a uniform magnetic field along [001] direction~\cite{2016PhysRevB.93.241202}.
Consequently, the triply degenerate nodal points along $\Gamma$-A are absent.
A deeper inspection reveals that there are four Weyl points in the whole Brillouin zone (BZ) (See SM~\cite{SM} for details).
These four Weyl points located along the rotation-invariant high-symmetry line $\Gamma$-A are constrained by the rotation symmetry $\mathcal{C}_3$, which indicates that CPL irradiation breaks $\mathcal{\sigma}_{v}\oplus\mathcal{T}$ but preserve $\mathcal{C}_3$.
To illustrate the nontrivial band topology of the CPL-induced Weyl semimetal phase in WC-type TiO, we present the surface states of  CPL-irradiated WC-type TiO.
The obtained local density of states (LDOS) projected on the semi-infinite (010) surface are plotted in Figs.~\ref{fig:fig3}(d) and ~\ref{fig:fig3}(e), and one can find the characteristic surface states terminated at projections of bulk Weyl nodes.

\begin{figure}[t]
\includegraphics[width=\linewidth]{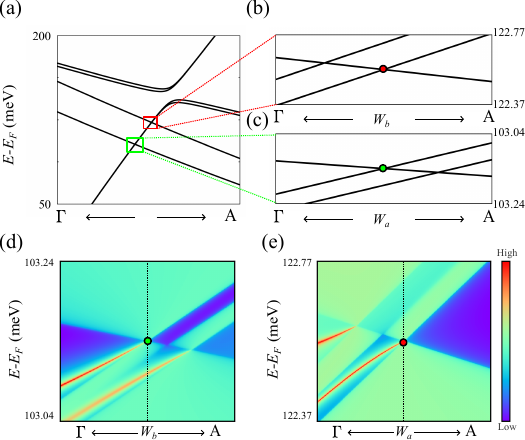}
\caption{\label{fig:fig2}
WC-type TiO under 4\% uniaxial tensile strain.
(a)-(b) Schematic figure of WPs location in BZ, with red and green dots represent the Weyl nodes with chirality equal to $+1$ and $-1$, respectively.
(c) Band structure along $\Gamma$-A reveal a gap opening along hight-symmetry lines.
(d)-(e) The surface spectrum of (010) surfaces along the high-symmetry line connected Weyl nodes with opposite chiralities and shown by the white dashed line in plot (f).
(f) Isoenergy band contours of semi-infinite (010) surfaces at 12.8 meV relative to the Fermi level.
}
\end{figure}

\textit{\textcolor{blue}{Floquet engineering of strained WC-type TiO and flexible controllable Weyl semimetallic phase. ---}}
Although CPL-irradiated WC-type TiO gives rise to a Weyl semimetallic phase, the band splitting of $\Lambda _{6}$ is too small to allow the detection of Fermi arcs.
To achieve an experimentally feasible Weyl semimetal phase, it is necessary to break additional symmetries.
In the following, we will see that the application of lattice strain and its coupling of CPL would offer an effective way to manipulate the electronic and topological properties of WC-type TiO.
To be specific, we here first verify the evolution of band topology by applying lattice strain that breaks the rotational symmetry $\mathcal{C}_3$.
The band structure of WC-type TiO under 4\% tensile strain with SOC is shown in Fig.S2 \cite{SM}, and it is found that the local band gap is always present alone the high-symmetry lines.
Even so, two inequivalent Weyl nodes are found in $k_y=0$ plane in momentum space.
Due to the presence of $\mathcal{\sigma}_{h}$ and $\mathcal{T}$, there are eight Weyl nodes in the whole BZ, as schematically shown in Figs.~\ref{fig:fig2}(a) and \ref{fig:fig2}(b).
We tabulate the coordinates of the two inequivalent Weyl nodes labeled as $W_1$ and $W_3$ in Table ~\ref{tab:table_both}.
The obtained LDOS projected on the semi-infinite (010) surface is presented in Figs.~\ref{fig:fig2}(d) and \ref{fig:fig2}(e), where the surface states connecting the projections of bulk Weyl nodes are clearly visible.
To exhibit the appearance of Fermi arcs, we further examine isoenergy contours of Fermi surface projected on the semi-infinite (010) surface.
The isoenergy contours of (010) surfaces at 12.8 meV relative to the Fermi level is presented in Fig.~\ref{fig:fig2}(f).
We can see that, there are four open Fermi arcs in the (010) Fermi surface.
Interestingly, the Fermi arcs can be divided into two sets that are nearly parallel to each other.
These unique Fermi arcs provide multiple surface charge transport channels.
Therefore, in WC-type TiO, remarkable quantum phenomena originating from characteristic topological Fermi arcs, such as the 3D quantum Hall effect~\cite{tang2019three}, are expected to be different from previous works.

\begin{table}[b]
    \caption{\label{tab:table_both}
    Weyl nodes coordinates in moment space for intrinsic 4\% strained WC-type TiO under CPL with intensity of $e \mathrm{A}_0/\hbar = 0.000$ \AA $^{-1}$, $e \mathrm{A}_0/\hbar = 0.050$ \AA $^{-1}$  and $e \mathrm{A}_0/\hbar = 0.100$ \AA $^{-1}$. 
    Here, we just tabulate the non-equivalent Weyl nodes $W_1$ and $W_3$ with chirality $+1$. Due to the $\mathcal{\sigma}_{h}$ and $\mathcal{T}$ symmetries, we can get the positions of the other six Weyl nodes.
     }
    \begin{ruledtabular}
    \vspace{0.150cm}
\renewcommand\arraystretch{1.3}
    \begin{tabular}{cccccccc}
      Light intensity &  & $C$ & $k_x (\mathrm{\AA^{-1}})$ & $k_y (\mathrm{\AA^{-1}})$ & $ k_z (\mathrm{\AA^{-1}})$ & energy(eV) \\
    \hline
    \multirow{2}{*}{0.000 \AA $^{-1}$}
    &$W_1$ &+1&  0.0653  &  0.0000  & -0.7331&0.0128  \\
    &$W_3$ &+1&  0.1204  &  0.0000  & -0.7767&0.0126  \\
    \hline
    \multirow{2}{*}{0.050 \AA $^{-1}$}
    &$W_1$ &+1& 0.0564 & 0.0000 & -0.7296 & 0.0212 \\
    &$W_3$ &+1& 0.1132 & 0.0000 & -0.7686 & 0.0185 \\
    \hline
    0.100 \AA $^{-1}$ &$W_3$ &+1& 0.0990 & 0.0000 & -0.7547 & 0.0241 \\
    \end{tabular}
    \end{ruledtabular}
\end{table}

\begin{figure}[t]
\includegraphics[width=0.86\linewidth]{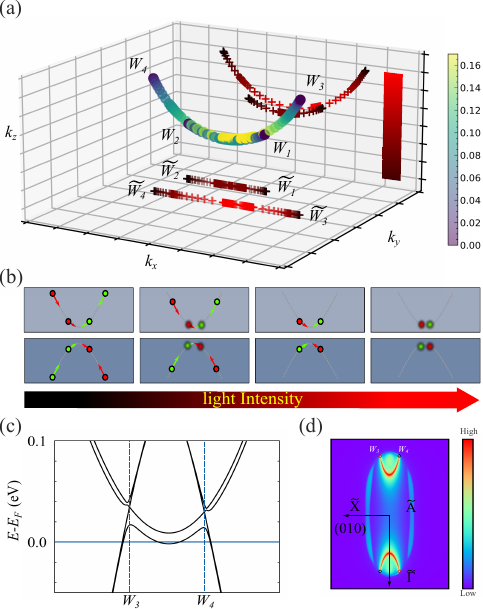}
\caption{\label{fig:fig4}
Floquet engineering on WC-type TiO under CPL coupled with  4\% tensile strained .
(a) The dynamic behaviors of WPs under CPL with different intensity. The colored circles mark the position of WP, and the colored `+' sign marks the projection of those location. The color bar denotes the values
of light intensity \AA$^{-1}$ . The letters marked the (projected) location of each Weyl points under no CPL light field.
(b) Schematic figure of dynamic behavior of  Weyl points corresponding to (a).  The Weyl points with chirality of $+1/-1$ are marked in red/green dots. With the increase of light intensity, the Weyl points come closer in pairs and annihilate finally. And the arrows denote such movement in moment space
(c) The band dispersion around $W_3$ and $W_4$ under CPL with intensity of $ eA_0/\hbar = 0.100$ \AA $^{-1}$ .
(d) The isoenergy band contours of semi-infinite (010) surfaces under CPL with intensity of $ eA_0/\hbar = 0.100$ \AA $^{-1}$.
}
\end{figure}

To expand the manipulating degrees of freedom in WC-type TiO, we next couple CPL with lattice strain and carry out Floquet engineering on strained WC-type TiO.
With various intensities of CPL, the strained WC-type TiO exhibits distinct Weyl semimetallic phases with different numbers of Weyl nodes and Fermi arcs.
As depicted in Fig. S5 \cite{SM} and tabulated in Table~\ref{tab:table_both}, as the light intensity increases, the photon-dressed band structures of strained WC-type TiO show that the two lowest conduction bands and the two highest valence bands simultaneously move away from the Fermi level.
Consequently, the Weyl nodes $W_1$ and $W_2$ come together and then annihilate at a specific light intensity of $e \mathrm{A}_0/\hbar = 0.089$ \AA $^{-1}$.
To deeply investigate the evolved behaviors of Weyl nodes, the positions of Weyl nodes as a function of light intensity are shown in Figs.~\ref{fig:fig4}(a) and \ref{fig:fig4}(b), and the positions of Weyl nodes with a light intensity of
$e \mathrm{A}_0/\hbar = 0.000$, $0.050$ and $0.100$ \AA $^{-1}$ are listed in Table~\ref{tab:table_both}.
We find that the light-driven motion of Weyl nodes is restricted to the $k_y=0$ plane.
When the light intensity is greater than $e \mathrm{A}_0/\hbar = 0.170$ \AA $^{-1}$, we notice that all Weyl nodes are annihilated and the strained WC-type TiO turns to a trivial insulator.
To reveal light-dressed topological features of strained WC-type TiO, we here present the band structures with $e \mathrm{A}_0/\hbar = 0.100$ \AA $^{-1}$ along the high-symmetry line connected Weyl nodes $W_3$ and $W_4$ in Fig.~\ref{fig:fig4}(c).
In this case, there are four Weyl nodes in the whole BZ, and the isoenergy contours of Fermi surface projected on (010) surface at 24.1 meV relative to the Fermi level are ploted in Fig.~\ref{fig:fig4}(d), where the Fermi arcs connect the projections with bulk Weyl nodes with opposite chirality are clearly shown.
In short, by carrying out Floquet engineering of strained WC-type TiO, the coupling of CPL with lattice strain can lead to intriguing Weyl semimetallic phases, where the number and position of Weyl nodes and Fermi arcs are highly controllable.

\textit{\textcolor{blue}{Summary. ---}}
In summary, by the means of first-principle calculations, Floquet theorem, and symmetry analysis, we carried out CPL Floquet engineering of topological triple fermions in strain-free and strained WC-type TiO.
We found that the symmetry-breaking perturbations such as periodic-driving CPL and lattice strain can induce specific symmetry breaking.
Consequently, this leads to destroy the triply degenerate nodal points and subsequently drive the system towards abundant Weyl semimetallic phases.
Constrained by the survived spatial symmetry, the Weyl nodes in WC-type TiO are demonstrated to be located in the high-symmetry plane or along the high-symmetry line in momentum space.
More importantly, the coupling of CPL with lattice strain can generate intriguing Weyl semimetallic phases with highly controllable Weyl nodes and Fermi arcs.
Besides CPL and strain, perturbations such as square wave that break $\mathcal{C}_3$ or $\mathcal{\sigma}_{v}\oplus\mathcal{T}$ could drive TiO to Weyl semimetal.
In the third section of SM~\cite{SM}, taking the square wave as an example, the triple fermion is verified to be driven in to Weyl fermion. 
Considering that the topological triple fermions in materials with WC-type structure (e.g., TiO, MoP, WC, TaN, ZrTe)  have been well-studied \cite{2016PhysRevX.6.031003,2016PhysRevB.93.241202,lv2017observation,ma2018three,ullah2023first,jin2023strain}, our results would facilitate the profound interest in theories and experiments.
Besides, as light-driving TSMs with triply degenerate nodal points  currently remains unexplored, our findings are expected to advance Floquet engineering of CPL in investigating topological triple fermions in realistic condensed matter materials.

\textit{\textcolor{blue}{Acknowledgments. ---}}
This work was supported by the National Natural Science Foundation of China (NSFC, Grants No.~12204074, No.~12222402, No.~92365101, No.~12347101, and No.~12074108) and the Natural Science Foundation of Chongqing (Grant No.~2023NSCQ-JQX0024, No.~CSTB2022NSCQ-MSX0568). D.-S M. also acknowledges funding from the China National Postdoctoral Program for Innovative Talent (Grant No.~BX20220367) and the Project Funded by China Postdoctoral Science Foundation (Grants No.~2023M740411).

\bibliography{Manuscript}

\end{document}